\documentclass[aps,pre,showpacs,floatfix,preprint]{revtex4-1}
\usepackage{graphicx}
\usepackage{dcolumn}
\usepackage{bm}
\usepackage{amsmath,amssymb}

\begin{document}

\title{Molecular-dynamics simulation of clustering processes in
sea-ice floes}

\author{Agnieszka Herman}
\email{oceagah@univ.gda.pl}
\homepage{http://www.ocean.univ.gda.pl/~herman/}
\affiliation{Institute of Oceanography, University of Gda\'nsk,
81-378 Gdynia, Poland}

\begin{abstract}
In seasonally ice-covered seas and along the margins of perennial ice
pack, i.e. in regions with medium ice concentrations, the ice cover
typically consists of separate floes interacting with each other by
inelastic collisions. In this paper, hitherto unexplored analogies
between this type of ice cover and two-dimensional granular gases are
used to formulate a model of ice dynamics at the floe
level. The model consists of: (i) momentum equations for floe motion
between collisions, formulated in the form of a Stokes-flow problem,
with floe-size dependent time constant and equilibrium velocity, and
(ii) hard-disk collision model. The numerical algorithm developed is
suitable for simulating particle-laden flow of $N$ disk-shaped floes with arbitrary size distribution. The model is applied
to study clustering phenomena in sea ice with power-law floe-size distribution. In particular, the influence of the average ice concentration $\bar{A}$ on the formation and characteristics of clusters is analyzed in detail. The results show the existence of two regimes, at low and high ice concentration, differing in terms of the exponents of the cluster-size distribution and of the size of the largest cluster.
\end{abstract}


\maketitle

\section{Introduction}\label{sec:intro}

Recent global climate change, amplified in the polar regions of the
earth, has led during the last decades to a substantial reduction of
the ice extent, especially in the Arctic \cite{Serrezeetal07}. This
trend is strongly coupled to a decrease of the surface area covered
with thick, multi-year ice, and a corresponding expansion of thin,
seasonal ice-cover type \cite{Maslaniketal07,Rothrocketal08}. Whether
the thinning of the ice cover will eventually lead to permanently
ice-free summers in the Arctic is still a subject of debate
\cite{Serreze11}. Undoubtedly, however, it substantially changes the
response of the ice cover to the external forcing. Among other
things, lower mechanical strength of thin ice makes it more
susceptible to deformation, breaking and divergence. This kind of
highly mobile ice cover, until recently occurring along the edges of
the perennial ice pack (a so-called marginal ice zone, MIZ) and in
subpolar, seasonally ice-covered seas, has been incomparably less
intensively studied than the central Arctic ice pack. Global climate
trends suggest that expanding our knowledge of the dynamics of this
type of ice cover will become increasingly important in the future.

In this paper, a term `medium-concentration ice zone' (MCIZ) is
used rather broadly to describe a strongly fragmented ice cover type,
consisting of clearly separated floes, with ice concentrations
$\bar{A}<1$, i.e., in a dynamic regime between a free drift
($\bar{A}\ll1$) and a compact ice cover ($\bar{A}\simeq1$). Within
MCIZ defined in this way, sea ice possesses a number of properties
characteristic of granular materials -- assemblies of discrete,
macroscopic solid particles, interacting with each other via
dissipative forces (e.g., friction and/or inelastic collisions).
Dissipative nature of interactions is responsible for distinctive
properties of granular materials, different from properties of
fluids, solids and gases. For a review of collective properties and
pattern formation in granular materials see, e.g., Aranson and
Tsimring \cite{AransonTsimring06}.

Sea ice in MCIZ has all properties required to regard it as a
particular subclass of granular materials: a two-dimensional (2D)
granular gas. It consists of discrete solid floes (particles)
immersed in water/air (`interstitial fluids') and moving on the sea
surface due to the action of external forces of wind, currents etc.
The floes can be regarded as rigid and indeformable. They
collide inelastically, i.e., with a loss of kinetic energy. At medium
ice concentrations floe--floe collisions contribute substantially to
the internal stress in the ice \cite{Shen84,Shen86,Lu89,Feltham05}.
Equations for the slowly-varying and random component of the floes'
velocity \cite{Feltham05} are fully analogous to those for
dissipative gases \cite{GoldhirschZanetti93,BreyCubero01}. To the
best of my knowledge, those analogies have never been explored in
sea-ice modeling beyond the study by Feltham \cite{Feltham05} who
applied granular flow theory to formulate a simplified MCIZ model
with a composite collisional and plastic rheology, and showed that
narrow zones of rapid ice movement along the MIZ edge (ice jets) can
be obtained as an analytical solution to this model. It has been long
recognized that collisional rheology is very different from the
plastic one, routinely used in sea-ice models. However, the available
collisional rheology models \cite{Shen84,Shen86,Lu89}, including the
one used by Feltham, are based on unrealistic assumptions of floes of
equal size or very narrow size distributions, uniformly distributed
on the sea surface.

\begin{figure*}[!t]
  \noindent\includegraphics[width=40pc]{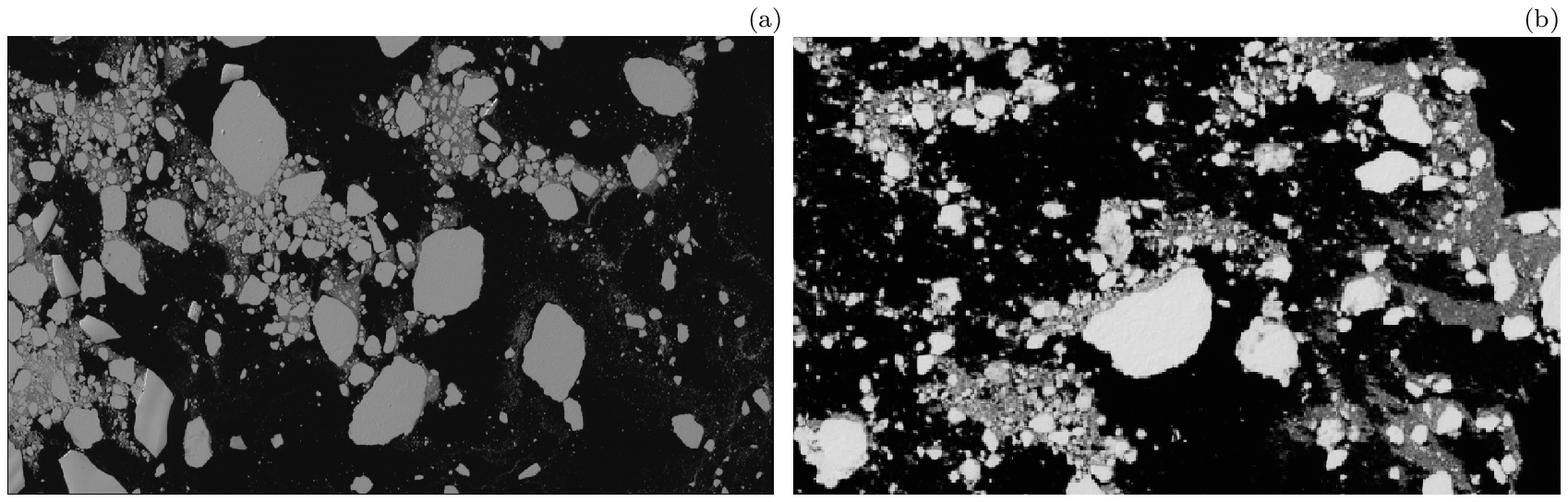}\\
  \caption{Fragments of Landsat images showing clustered sea-ice floes
  off the Antarctic coast (at 121.8$^\circ$W, 73.5$^\circ$S)
  on 18. Feb. 2008 (a) and 26. Jan. 2008 (b). Source: \cite{LandsatData}.}\label{fig:examples}
\end{figure*}

With all above-mentioned analogies between MCIZ and granular-gas
dynamics in mind, there are a number of features that make sea ice
unique among typically studied granular-gas problems. The most
important of those features are, firstly, very wide, scale-free size
distributions of ice floes, and secondly, size-dependent equilibrium
velocities of floes under a given external forcing. The floe-size
distributions (FSDs) observed in many ocean regions (see
\cite{Inoue04,Toyota06,Steer08,Herman10,Toyota11} for recent works),
are of power-law type $P(d)\sim d^{-1-\alpha_r}$, where $d$ denotes the
floe diameter and $\alpha_r>1$. Most recent studies, based on combined satellite and
airborne data covering a very wide range of floe sizes, showed the
existence of two `regimes' in FSDs for large and small floes,
respectively, although the nature of the transition between those
regimes has not been established yet
\cite{Toyota06,Steer08,Herman10,Toyota11}. In this work, we assume
for simplicity that $\alpha_r$ is a constant, and concentrate on the
influence of the heavy FSD tail on the patterns of motion and
cluster formation in MCIZ.

This study draws from the analogies between sea ice in MCIZ and
granular gases, with a goal to develop a theoretical and numerical
model of MCIZ dynamics at the floe level, and to explore the distinct
behavior of sea ice compared to other, well studied granular-gas
systems. Starting from well established equations of sea-ice motion,
it is demonstrated that, under certain assumptions, the momentum
equation describing the floe's response to wind and current can be
formulated as a Stokes-flow problem, with a floe-size dependent time
constant and equilibrium velocity. Combined with a hard-disk (HD)
model of floe--floe collisions, the Stokes equations are used to
formulate a numerical model of MCIZ dynamics, in which, contrary to
previous attempts, floe--floe interactions are simulated directly. In
the further parts of this paper, the model is applied to study one of
the fundamental aspects of granular-gas dynamics, namely formation of
clusters due to irreversible loss of kinetic energy of colliding
inelastic particles (see~\ref{sec:theory} below). The general idea is
that clusters of floes, observed in many real situations
(Fig.~\ref{fig:examples}; see also \cite{Toyota11}), can be explained
as an inherent property of sea ice, and not necessarily as a
`passive' result of the external forcing (convergence of surface
currents).

A number of possible further applications of the model developed here
are beyond the scope of this paper and will be explored elsewhere.
They include formulation of internal-stress parametrization schemes
suitable for MCIZ (at present, as mentioned above, only very crude
collision-rheology models are available and in large-scale sea-ice
models the internal stress is taken into account only at
$\bar{A}\simeq1$, see, e.g., \cite{CICE41}), and improved formulae
for the atmosphere-ocean momentum and heat transfer in MCIZ. Some possible consequences of cluster formation in sea ice for the above processes,
as well as the existence of possible feedbacks, are discussed at the
end of this paper.

Apart from sea-ice modeling applications, the results of this work
may be also of interest to researchers studying behavior of granular
gases (particle-laden flows in particular) with wide particle-size
distributions, which are still relatively unexplored theoretically
and numerically \cite{Luding09}. In particular, hitherto studies of
clustering processes in polydisperse granular media concentrated
either on binary or narrow size distributions
\cite{LudingStrauss01,DahlClellandHrenya2002,RiceHrenya2010}. The
model and code developed here can be easily applied to other
granular-gas systems consisting of particles with any given size
distribution and size-dependent response to forcing.

The structure of this paper is as follows: further parts of the
Introduction provide a brief theoretical background of cluster
formation in granular gases, measures of cluster properties and
methods of molecular-dynamics (MD) modeling of granular gases.
Readers familiar with physics and modeling of granular materials may
skip this part. In Section~\ref{sec:equations}, the equations of
ice-floe motion are formulated in a way suitable for MD simulation.
The model and its numerical implementation are described in
Section~\ref{sec:model}, followed by the presentation of the modeling
results in Section~\ref{sec:results}. The paper concludes with a discussion in Section~\ref{sec:conclusions}.

\subsection{Clustering in granular gases -- mechanisms and
cluster characteristics }\label{sec:theory}

In granular gases, clusters -- regions with higher than average
particle density -- are initiated by random density fluctuations.
Increased collision frequency in regions with higher density leads to
a local decrease of the `granular temperature' (kinetic energy
associated with the random component of the particle motion,
analogous to the thermodynamic temperature of gases), and thus to a
local decrease in pressure. As a result, clusters tend to grow by
attracting particles from their neighborhood, so that they may
persist for a long time \cite{GoldhirschZanetti93}. This phenomenon
is known from real dissipative granular systems and has been
extensively studied numerically (e.g.,
\cite{LudingHerrmann99,DahlClellandHrenya2002,Luding02,MillerLuding04,Luding05,RiceHrenya2009,RiceHrenya2010}).
The tendency to form clusters, as well as their characteristic size,
depend on the average properties of the system (particle density,
coefficient of restitution, particle size and mass distribution,
etc.), as well as on the external forcing. Most studies on granular
materials consider equal-size and equal-mass particles. In the case
of continuous size distributions, larger particles tend to accumulate
in the central regions of clusters (a `self-sorting mechanism',
\cite{RiceHrenya2010}).

A number of parameters can be defined for the purpose of a
quantitative description of clustering processes. The most
straightforward way of defining clustered/diluted regions is in terms
of the local particle concentration $A$ relative to the average
concentration $\bar{A}$ in the analyzed region (e.g.,
\cite{RiceHrenya2009,RiceHrenya2010}). In this paper, another,
distance-based approach is taken, i.e., the particles are assigned to
clusters in such a way that every two particles separated by a distance
smaller that $d_m$ belong to the same cluster (e.g.,
\cite{MillerLuding04,Luding05}). Cluster properties are defined as averages and/or distributions of the
properties of particles forming them, see Section~\ref{sec:results}.
Additionally, a frequently used global measure of the spatial
particle distribution is a radial distribution function
$g_{rdf}(x)$, describing the expected particle concentration at a distance $x$ from the center of a randomly selected particle, normalized with $\bar{A}$ \cite{LudingStrauss01,RiceHrenya2009,RiceHrenya2010}. This function is particularly suitable for monodisperse systems \cite{LudingStrauss01}, as it can provide a measure of characteristic cluster size. For polydisperse systems, floe--floe distance distribution $g_{ffd}(x)$ proposed in Section~\ref{sec:resultsB} seems more suitable.

\subsection{Modeling of granular gases}\label{sec:ggmodeling}

At the core of an MD simulation is the detection of time and partners of
collisions among a (usually very large) number of particles constituting the analyzed system. Hence, algorithms for solving typical
fluid-dynamics problems, in which the time is advanced by prescribed,
constant time steps, are unsuitable. Instead, event-driven
algorithms (EDA) must be used
\cite{AllenTildesley89,Sigurgeirsson01}. The choice of a proper EDA
depends on (i) the character of motion of the particles between
collisions, i.e., their interactions with the surrounding fluid, and (ii) the nature of particle--particle collisions. Most
relevant for this work are algorithms designed for a subclass of
dissipative-dynamics problems called particle-laden flows
\cite{Sigurgeirsson01}, in which the motion of the fluid is given, i.e.,
it influences, but is not influenced by particles immersed in it.

Considering the material properties of sea ice it is justified to use a hard-disk (HD) collision model, assuming that at collision the momentum is transferred instantaneously along the line joining the centers of colliding particles \citep{AllenTildesley89}. In other words, collisions are pairwise and infinitely short. EDAs for HD models are built based on event lists, in which times of collisions between all pairs of (potentially colliding) particles are stored and updated every time a collision takes place or particles' velocities change. Available EDAs differ mainly in terms of how the event lists are updated, which may have a profound influence on the computational efficiency of those EDAs. For examples, see \cite{Marin97,MillerLuding03} and a review by \cite{Sigurgeirsson01}.

\section{Assumptions and equations}\label{sec:equations}

Let us consider a set of $N$ disk-shaped, non-overlapping sea-ice
floes with diameters $d_i$ ($i=1,\dots,N$), equal heights $h$ and
constant density $\rho$, so that their mass centers coincide with
their geometric centers. Due to the action of external forces, the
floes move on the sea surface along trajectories $\mathbf{x}_i(t)$,
where $\mathbf{x}_i\in\mathcal{A}$ denotes the position vector within
an analyzed region $\mathcal{A}\subset\mathbb{R}^2$ and $t$ denotes
time. The state of each floe at time $t$ is given by
$(\mathbf{x}_i(t),\mathbf{u}_i(t))$, where
$\mathbf{u}_i=\mathrm{d}\mathbf{x}_i/\mathrm{d}t$ denotes the floe's
translational velocity. Without interactions with neighboring floes,
each floe moves independently and its acceleration can be determined
from its state and the external forces, as described below. If the
trajectories of floes intersect, they collide inelastically, with a loss of kinetic energy.

\subsection{Floe motion between collisions}

\begin{figure}
  \noindent\includegraphics[width=20pc]{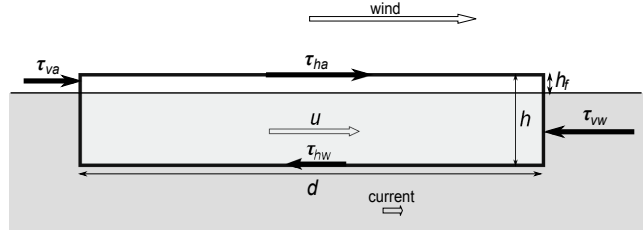}\\
  \caption{A side view of a disk-shaped ice floe under the action of
  wind and current. The arrows are not to scale. See the text for explanation of the symbols.}\label{fig:forces}
\end{figure}

Between collisions, the motion of the $i$-th floe satisfies the
momentum conservation equation (e.g., \cite{LepparantaBook}):
\begin{equation}\label{eq:momentum}
    m_i\left({\mathrm{d}\mathbf{u}_i\over\mathrm{d}t} +
    f\mathbf{k}\times\mathbf{u}_i\right) =
    \int_{V_i}\mathbf{F}_{b,i}\mathrm{d}V +
    \int_{S_i}\mathbf{F}_{s,i}\mathrm{d}S,
\end{equation}
where $\mathbf{k}=[0,0,1]$, $f$ denotes the Coriolis parameter,
$m_i=\rho hS_i$ is the floe mass, $S_i=\pi d_i^2/4$ -- its
upper/lower surface area, and $\mathbf{F}_{b,i}$, $\mathbf{F}_{s,i}$
denote the sum of body and surface forces, respectively, acting on
the analyzed floe. For simplicity, we will further assume that $f=0$
and $\mathbf{F}_{b,i}=0$ (in particular, the force resulting from the
gradient of the geopotential height of the sea surface is not taken
into account). The net surface force $\mathbf{F}_{s,i}$ results from
a sum of four terms (Fig.~\ref{fig:forces}): stress acting on the
upper and lower surface of the floe (atmospheric and oceanic skin
drag, $\boldsymbol\tau_{ha,i}$ and $\boldsymbol\tau_{hw,i}$,
respectively), and pressure acting on the floe's edges
(atmospheric and oceanic body drag, $\boldsymbol\tau_{va,i}$ and
$\boldsymbol\tau_{vw,i}$, respectively), see, e.g.,
\cite{MaiWamserKottmeier96,Garbrechtetal02,LupkesBirnbaum05}. The
vertical surface area exposed to $\boldsymbol\tau_{va,i}$ and
$\boldsymbol\tau_{vw,i}$ equals $d_ih_f$ and $d_i(h-h_f)$,
respectively, where $h_f=h(\rho_w-\rho)/\rho_w$ denotes the floe's
freeboard and $\rho_w$ denotes water density. (More realistically,
the surface areas depend on the height/depth of ridges/keels,
respectively, but this effect is not taken into account here.) Thus,
(\ref{eq:momentum})~can be rewritten as:
\begin{equation}\label{eq:momentum2}
    m_i{\mathrm{d}\mathbf{u}_i\over\mathrm{d}t} =
    S_i(\boldsymbol\tau_{ha,i}+\boldsymbol\tau_{hw,i}) +
    h_fd_i\boldsymbol\tau_{va,i} + (h-h_f)d_i\boldsymbol\tau_{vw,i}.
\end{equation}
In the following, the four forcing terms are parameterized with
simple drag formulae typically used in sea-ice and ocean modeling.
The atmospheric forcing depends quadratically on the wind speed
$\mathbf{u}_a$:
\begin{equation}\label{eq:taua}
    \boldsymbol\tau_{ha,i} = \rho_aC_{ha}|\mathbf{u}_a|\mathbf{u}_a,
    \quad
    \boldsymbol\tau_{va,i} = \rho_aC_{va}|\mathbf{u}_a|\mathbf{u}_a,
\end{equation}
where $\rho_a$ denotes the air density. The oceanic forcing depends
on the floe velocity relative to the surface current $\mathbf{u}_w$.
A quadratic relationship analogous to~(\ref{eq:taua}) is replaced
with a linear one \cite{LepparantaBook}:
\begin{equation}\label{eq:tauw}
    \boldsymbol\tau_{hw,i} = \rho_wC_{hw}(\mathbf{u}_w-\mathbf{u}_i),
    \quad
    \boldsymbol\tau_{vw,i} = \rho_wC_{vw}(\mathbf{u}_w-\mathbf{u}_i).
\end{equation}

\begin{figure}[!t]
  \noindent\includegraphics[width=20pc]{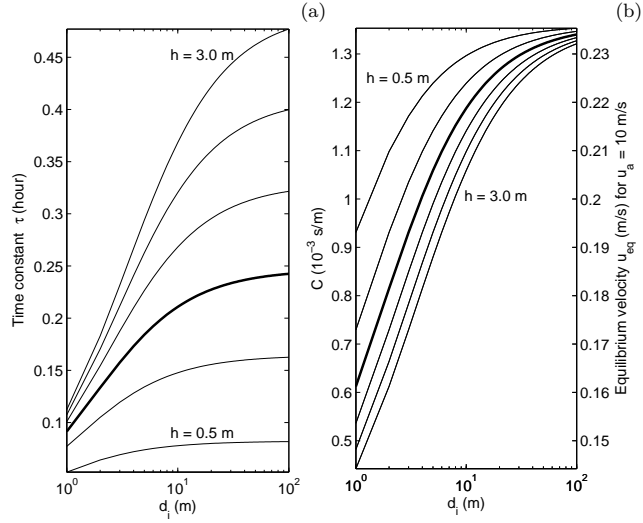}\\
  \caption{Time constant $\tau$ (a), coefficient $C$ and the
  equilibrium floe velocity $|\mathbf{u}_{eq}|$ for wind speed
  $|\mathbf{u}_a|=10$~m/s (b) in function of the floe diameter $d_i$,
  for a set of ice thickness values $h$. The lines are drawn every
  0.5~m from $h=0.5$~m to $h=3.0$~m. Thick lines correspond to
  the model setup used in the simulations described in
  section~\ref{sec:results} (see Table~\ref{tab:params}).}\label{fig:params}
\end{figure}

The drag coefficients $C_{ha}$, $C_{va}$, $C_{hw}$, and $C_{vw}$ are
assumed constant. Linearization of $\boldsymbol\tau_{hw,i}$ and
$\boldsymbol\tau_{vw,i}$ has important consequences for numerical
algorithms used in the model, as it enables to formulate the
governing equation~(\ref{eq:momentum2}) in the form mathematically
analogous to the Stokes-flow problem \cite{Sigurgeirsson01}. Namely,
from~(\ref{eq:momentum2})--(\ref{eq:tauw}) we have:
\begin{equation}\label{eq:momentumfinal}
    \tau_i{\mathrm{d}\mathbf{u}_i\over\mathrm{d}t} =
    \mathbf{u}_{eq,i} - \mathbf{u}_i,
\end{equation}
where the time constant $\tau_i$ is given by:
\begin{equation}\label{eq:tau}
    \tau_i = \left[{\rho_wC_{hw}\over\rho h}+{4C_{vw}\over\pi d_i}\right]^{-1}.
\end{equation}
After enough time and without interactions with other floes, each
floe would reach its equilibrium
($\mathrm{d}\mathbf{u}_i/\mathrm{d}t=0$), free-drift velocity
$\mathbf{u}_{eq,i}$, given by:
\begin{equation}\label{eq:equilvel}
    \mathbf{u}_{eq,i} = \mathbf{u}_w + C_i|\mathbf{u}_a|\mathbf{u}_a,
\end{equation}
where:
\begin{equation}\label{eq:C}
    C_i = \tau_i\left[{\rho_aC_{ha}\over\rho h} +
    {4(\rho_w-\rho)\rho_aC_{va}\over\pi\rho_w\rho d_i}\right].
\end{equation}
As can be seen, due to the form-drag effects $\tau_i$ and
$\mathbf{u}_{eq,i}$ are floe-size dependent, as shown in
Fig.~\ref{fig:params} for a selected range of model parameters.

\subsection{Floe--floe collisions}

Two floes, $i$ and $j$, moving along intersecting trajectories,
collide at time $t_c$ which is the smallest positive root of the
equation expressing the contact criterion for those floes:
\begin{equation}\label{eq:collisiontime}
    ||\mathbf{x}_i(t_c)-\mathbf{x}_j(t_c)||=(d_i+d_j)/2.
\end{equation}
As already mentioned, it is assumed that collisions are inelastic and infinitely short, and
that they can be resolved via a HD model. If $\mathbf{u}_i$,
$\mathbf{u}_j$ denote velocities just before collision, the
post-collision velocities, $\tilde{\mathbf{u}}_i$ and
$\tilde{\mathbf{u}}_j$, are given by:
\begin{equation}\label{eq:collisionvel}
    \tilde{\mathbf{u}}_i = \mathbf{u}_i - m_j\mathbf{n}_{ij},\qquad
    \tilde{\mathbf{u}}_j = \mathbf{u}_j + m_i\mathbf{n}_{ij},
\end{equation}
where:
\begin{equation}
  \mathbf{n}_{ij} = {1+\epsilon\over m_i+m_j}
  \left(\mathbf{k}_{ij}\cdot(\mathbf{u}_i-\mathbf{u}_j)\right)\mathbf{k}_{ij},
\end{equation}
and $\mathbf{k}_{ij}$ is a unit vector pointing from $\mathbf{x}_i$
to $\mathbf{x}_j$. The restitution coefficient $\epsilon\in[0,1]$ is
assumed constant.

\section{The model}\label{sec:model}

Like other MD models (see Section~\ref{sec:ggmodeling}), the one developed in this study -- fancifully
called the Small Floe Collider (SFC) -- is based on an event-driven
algorithm, suitable for particle-laden flows. SFC is capable of
simulating a set of $N$ disc-shaped particles (`floes') with an
arbitrary size distribution, within a square domain of side length
$L$, with periodic boundaries. In order to simplify the treatment of
periodic boundaries, during initialization the model domain is scaled
to a region $[-1/2,1/2]\times[-1/2,1/2]$. All spatially-dependent
variables ($d_i$, $|\mathbf{u}_i|$ etc.) are scaled accordingly. In
order to speed up the simulations, the model domain is divided into
$N_c$ square computational cells in order to reduce the search region
for potential collision partners of a given floe. Consequently, two
event types must be handled: floe--floe collisions (FFC) and
virtual-wall collisions (VWC), corresponding to a transfer of a floe
to a neighboring cell \cite{Sigurgeirsson01}.

Before we describe the SFC algorithm, two issues are worth
discussing. Firstly, when discretizing the model equations, it should
be noted that the floes' trajectories between collisions can be
determined exactly from equation~(\ref{eq:momentumfinal}). With
initial conditions $\mathbf{u}_i(t_0)=\mathbf{u}_{i,0}$ and
$\mathbf{x}_i(t_0)=\mathbf{x}_{i,0}$, we obtain easily, for $\Delta
t=t-t_0$:
\begin{equation}\label{eq:trajectory}
    \mathbf{x}_i(t) = \mathbf{x}_{i,0} + \Delta t\mathbf{u}_{eq,i} -
    \tau_i(\mathbf{u}_{i,0}-\mathbf{u}_{eq,i})(e^{-\Delta t/\tau_i}-1).
\end{equation}
Nevertheless, after inserting the floe positions given
by~(\ref{eq:trajectory}) into the collision
criterion~(\ref{eq:collisiontime}), the resulting equation cannot be
solved explicitly for $t_c$. Hence, (\ref{eq:trajectory}) must be replaced
with an approximate formula, allowing for a consistent calculation of floes' positions and
collision times. In SFC, the following linearly implicit
integrator is used:
\begin{eqnarray}
  \mathbf{x}_i(t) &=& \mathbf{x}_{i,0} + \Delta t\mathbf{u}_{i,0}, \label{eq:numint_x}\\
  \mathbf{u}_i(t) &=& \mathbf{u}_{i,0} +
                      \Delta t/\tau_i(\mathbf{u}_{eq,i}-\mathbf{u}_i(t)).
                      \label{eq:numint_u}
\end{eqnarray}
It can be shown easily that using (\ref{eq:numint_x}) amounts to
replacing the exponential term in~(\ref{eq:trajectory}) with the
first two terms in its series expansion ($e^x\approx1+x$).
From~(\ref{eq:trajectory}) and~(\ref{eq:numint_x}), the error of
floes' positions estimation can be calculated, and thus the maximum
acceptable $\Delta t_{max}$ at which velocities must be updated. The error tends to be larger for smaller
floes (because of both smaller $\tau_i$ and, typically, larger
$|\mathbf{u}_i-\mathbf{u}_{eq,i}|$, see further
section~\ref{sec:results}).

The second issue requiring some comment is related to the typical
numerical problem of hard-sphere models: the inelastic collapse due
to diverging collision rates between densely packed, almost touching
particles. In SFC, it is accounted for by the method of Luding and
McNamara \cite{LudingMcNamara98}: if a given floe participates in
more than one floe--floe collision within a certain time period
$\Delta t_c$, corresponding to a collision duration, the restitution
coefficient $\epsilon$ for those collisions is set to 1. In the
simulations described in this paper, $\Delta t_c=10^{-1}$~s has been
established experimentally as an optimal value.

\begin{figure}[!t]
  \noindent\includegraphics[width=17pc,clip=true]{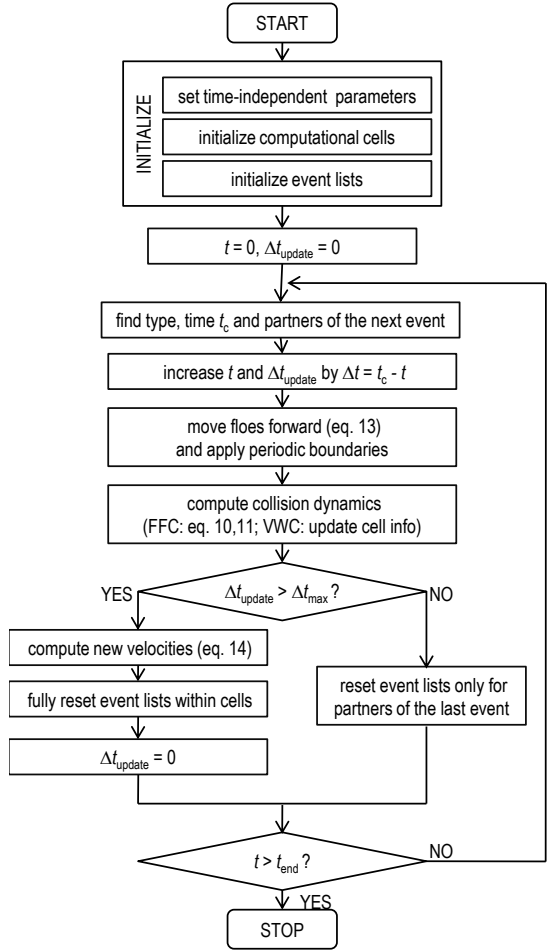}\\
  \caption{Flowchart of the SFC model.}\label{fig:flowchart}
\end{figure}

The main part of the computational algorithm
(Fig.~\ref{fig:flowchart}) can be summarized as follows: At every
iteration, the type of the nearest event, its partner(s) and time
$t_c$ are determined. Subsequently, the model time $t$ is advanced by
$\Delta t=t_c-t$, the floes are moved to their new positions, and the
event is handled, accordingly to its type (FFC or VWC). In the final part of the main loop, the event lists are either updated only for the partners of the last event, or fully recalculated for all particles -- depending on whether particle velocities have been updated at the current time step.

\section{Modeling results}\label{sec:results}

In this section, the results of SFC simulations will be presented, with emphasis on the influence of the ice concentration $\bar{A}$ on cluster formation. A detailed analysis of model behavior in the full range of parameters ($|\mathbf{u}_a|$, $\epsilon$, $\alpha_r$ etc.) is beyond the scope of this study. The general results in terms of floes' motion and cluster formation are described in Section~\ref{sec:resultsA}. A more detailed analysis of the cluster properties and their size distribution
is given in Sections~\ref{sec:resultsB} and ~\ref{sec:resultsC}.

The model configuration is summarized in Table~\ref{tab:params}. In
the simulations, the densities ($\rho$, $\rho_w$, $\rho_a$) and air-ice drag coefficients ($C_{ha}$, $C_{va}$) were set
to their typical values found in literature (e.g.,
\cite{LepparantaBook}). The linear ice-water drag coefficients were
estimated from their quadratic version ($\simeq3.5\cdot10^{-3}$) and
equations~(\ref{eq:tauw},\ref{eq:tau}--\ref{eq:C}) for
$|\mathbf{u}_i|\simeq|\mathbf{u}_{eq,i}|$, assuming $C_{ha}=C_{va}$
and $C_{hw}=C_{vw}$. The surface current speed relative to the wind
speed was estimated with the Ekman model \cite{LepparantaBook}. For
the ice floes properties, $h$, $\bar{d}$ and $\alpha_r$,
typical values observed in the Weddell Sea during the ISPOL
experiment \cite{Steer08} were chosen. The mean ice concentration $\bar{A}$ was varied between the subsequent model runs in order to investigate its influence on the modeling results.

\begin{table}
 \caption{Physical and numerical model parameters used in the SFC simulations\label{tab:params}}
 \begin{ruledtabular}
 \begin{tabular}{lccc}
 Parameter & Symbol & Value & Units \\
 \hline
 ice density                & $\rho$     & 910 &  kg/m$^3$ \\
 water density              & $\rho_w$   & 1025 & kg/m$^3$ \\
 air density                & $\rho_a$   & 1.23 & kg/m$^3$ \\
 air--ice skin-drag coef.   & $C_{ha}$   & $1.7\cdot10^{-3}$ & --- \\
 air--ice form-drag coef.   & $C_{va}$   & $1.7\cdot10^{-3}$ & --- \\
 water--ice skin-drag coef. & $C_{hw}$   & $1.3\cdot10^{-3}$ & m/s \\
 water--ice form-drag coef. & $C_{vw}$   & $1.3\cdot10^{-3}$ & m/s \\
 wind speed                 & $|\mathbf{u}_a|$ & 10 & m/s \\
 surface current speed      & $|\mathbf{u}_w|$ & $0.01|\mathbf{u}_a|$ & m/s \\
 ice thickness              & $h$        & 1.5 & m \\
 mean ice concentration     & $\bar{A}$  & 0.6--0.9 & --- \\
 restitution coef.          & $\epsilon$ & 0.85 & --- \\
 FSD slope                  & $\alpha_r$ & 1.5 & --- \\
 mean floe diameter         & $\bar{d}$  & 2.0 & m \\
 \hline
 No. floes                  & $N$        & 3000 & --- \\
 No. computational cells    & $N_c$      & $10^2$ & --- \\
 time step for vel. updates & $\Delta t_{max}$  & 1.0 & s \\
 collision duration         & $\Delta t_c$  & $10^{-1}$ & s \\
 No. ensemble model runs    & $N_{ens}$  & 5 & --- \\
 \end{tabular}
 \end{ruledtabular}
\end{table}

In all simulations, $\mathbf{u}_a=[u_a,0]$ and the number of floes $N=3000$. (Larger system sizes were tested for $\bar{A}=0.6$ and $\bar{A}=0.8$, but no differences in terms of statistical properties of the results were recorded.) The floe radii $d_i$, $i=1,\dots,N$, were obtained with maximum-likelihood estimation for a power-law distribution with a given exponent $\alpha_r$ (see appendix~\ref{sec:A_fsd} for details). In each case, the size of the model domain was adjusted to the generated set of the floe diameters, to obtain the desired average ice concentration $\bar{A}$. For each set of model parameters, $N_{ens}=5$ ensemble model runs were performed, differing in respect of initial conditions, i.e., the floes' initial velocities (drawn from a normal distribution with zero mean and standard deviation $1\cdot10^{-2}$~m/s) and random positions. The optimal number of computational cells $N_c$, the time step for updating the floes' velocity $\Delta t_{max}$, and the collision duration $\Delta t_c$ were established experimentally as $10^2$, 1.0~s and $0.1$~s,
respectively. For $\Delta t_{max}$ and $\Delta t_c$, the largest
values that had no significant influence on the modeling results were chosen.

In all cases, the model was run until a quasi-stationary state developed, characterized by insignificantly small temporal changes of such global variables as the total kinetic energy, collision rate and pressure. Comparisons between the ensemble model runs have shown that the initial conditions had no influence on the final values of those variables, although they may have influence on the details of the time evolution of the system towards the final state. The required number of iterations was established experimentally to $5\cdot10^6$ FFC events.

\subsection{General properties of the solution}\label{sec:resultsA}

\begin{figure}
  \noindent\includegraphics[width=20pc]{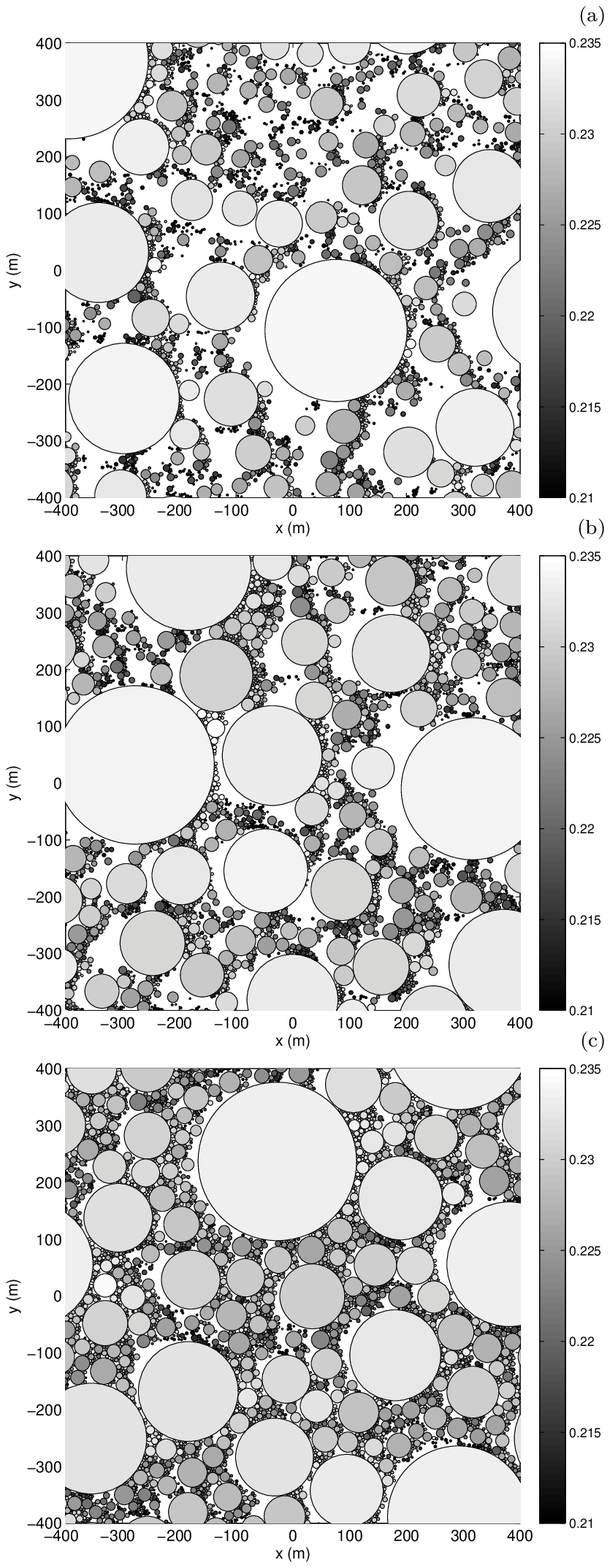}\\
  \caption{Snapshots (floe positions and velocities, in m/s, shown in gray scale) in a clustered state for $\bar{A}=0.6$~(a), $\bar{A}=0.725$~(b) and $\bar{A}=0.85$~(c). For better clarity, only equal-size fragments of the model domains are shown.}\label{fig:snapshots}
\end{figure}

\begin{figure*}
  \noindent\includegraphics[width=40pc]{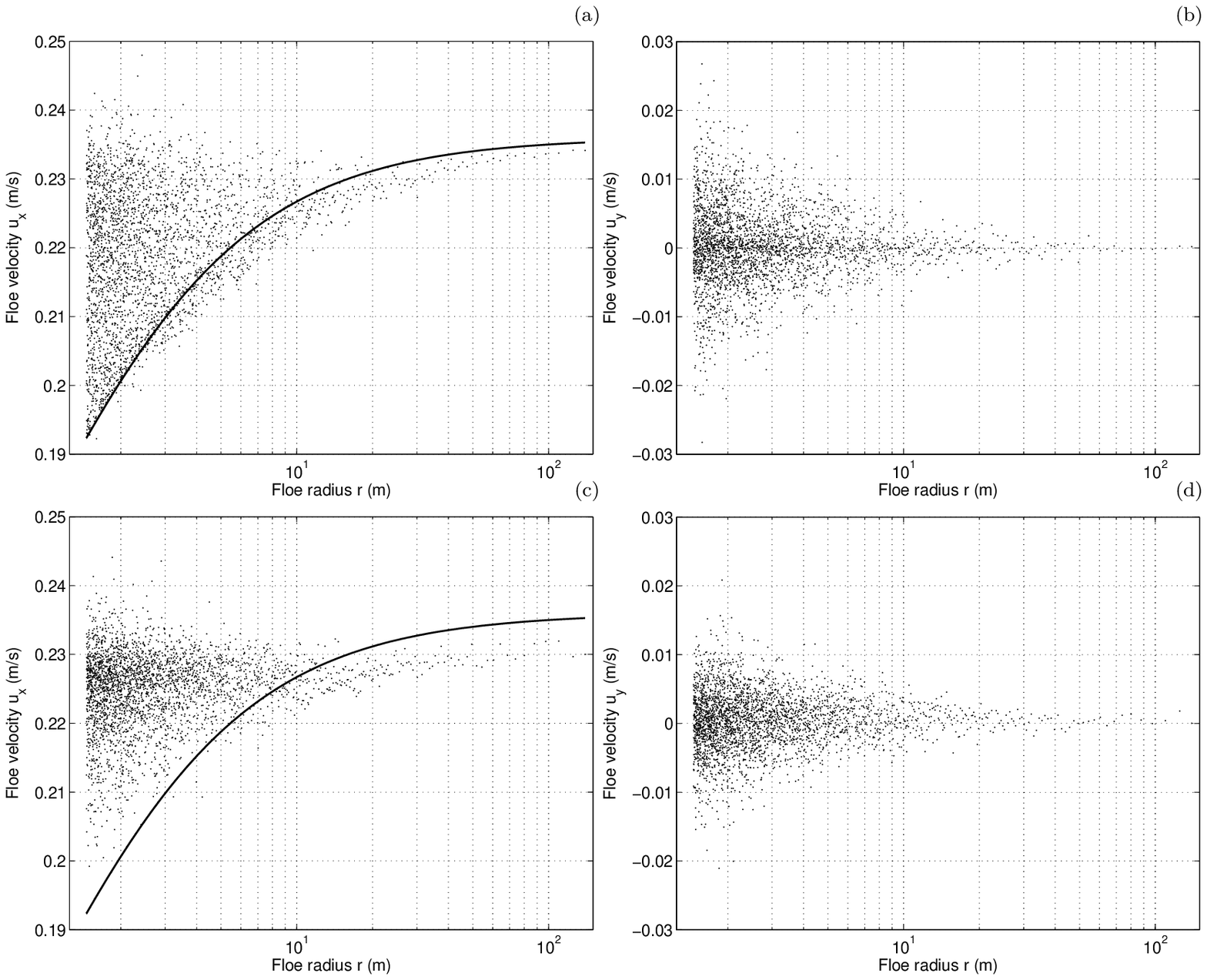}\\
  \caption{Scatterplots of instantaneous floe velocity components parallel  (a,c) and perpendicular (b,d) to the wind direction, as a function of the floe radius $r$, for $\bar{A}=0.6$~(a,b) and $\bar{A}=0.9$~(c,d). The lines in~(a) and (c) mark the equilibrium velocity
  $|\mathbf{u}_{eq}|$.}\label{fig:velcomps}
\end{figure*}

Figure~\ref{fig:snapshots} shows selected snapshots of the floes'
positions and velocities from three model runs with different $\bar{A}$. Contrary to the random initial state of the system, in the quasi-stationary regime the motion and distribution of the floes is patterned, with clearly developed clusters. Differently than in freely cooling granular gases, however, in the forced system under study the formation of clusters is strongly related to nonuniform equilibrium velocity of floes with different sizes. Larger floes tend to catch up with the smaller ones that accumulate in front of them and then slide along their sides, producing characteristic wakes. They are visible very well especially at small ice concentrations
(Fig.~\ref{fig:snapshots}a), but present even in densely packed ice-floe fields as narrow, elongated areas of open water upwind from large floes (Fig.~\ref{fig:snapshots}c). Within clusters, the floes tend to stay
in contact with each other. The translational velocity (i.e., velocity averaged over a certain number of collisions) is almost constant for all floes building a cluster. This translational motion is disturbed only if a cluster collides with another floe/group of floes. Unless such collision leads to a break-up of a cluster, velocities of floes constituting it
remain highly correlated. As a result, the along-wind velocity of the largest floes within clusters is typically lower than their equilibrium velocity (Fig.~\ref{fig:velcomps}a,c). This difference is larger for larger $\bar{A}$, i.e., higher collision rates. To the contrary, the smallest floes tend to move up to 30\% faster than their equilibrium velocity, with a random component of velocity in the order of 1--3$\cdot10^{-2}$~m/s, comparable to observed values, e.g. from the MIZEX experiment \cite{Shen84,Shen86}. Understandably, the higher the ice concentration, the lower the differences between the largest and the smallest floes (Fig.~\ref{fig:velcomps}c). Even though the differences $|\mathbf{u}_i-\mathbf{u}_{eq,i}|$ may be very large for individual floes, in all analyzed cases (i.e., independently of $\bar{A}$) the total momentum $\sum u_im_i$ develops in time towards $\sum u_{eq,i}m_i$, which corresponds to floe-mass-weighted average velocity of $\sim$0.23~m/s. At high ice concentrations, this value roughly corresponds to the highest density of points in the along-wind velocity scatterplot (and to the velocity of the largest floes, Fig.~\ref{fig:velcomps}c).

\subsection{Measures of clustering}\label{sec:resultsB}

\begin{figure*}
  \noindent\includegraphics[width=40pc]{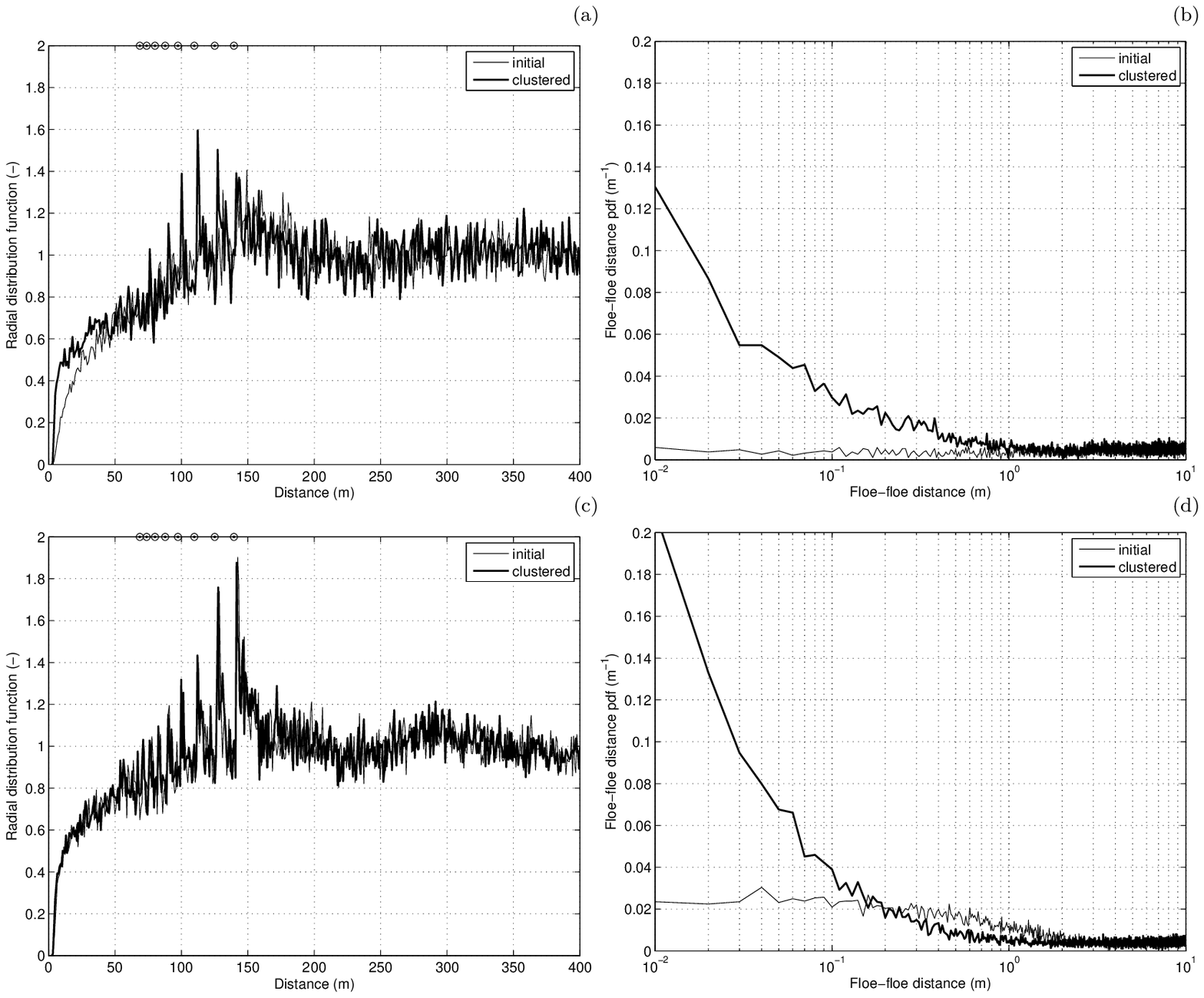}\\
  \caption{Radial distribution function $g_{rdf}$ (a,c) and floe--floe distance distribution $g_{ffd}$ (b,d) at the initial (random) and `final' (clustered) phases for $\bar{A}=0.6$~(a,b) and $\bar{A}=0.9$~(c,d). Symbols along the upper border of (a) and (c) show radii of the eight largest floes in the sample.}\label{fig:rdf}
\end{figure*}

Similarly as in observed clusters of ice floes (Fig.~\ref{fig:examples}), the behavior of the system is dominated by the largest floes (Fig.~\ref{fig:snapshots}). It is not surprising considering that among $N=3000$ power-law distributed floe diameters with $\alpha_r=1.5$, the largest floe occupies almost 10\% of the total floe area $F_{tot}=\bar{A}L^2=\pi/4\sum_{i=1}^N d_i^2$, and the ten largest floes occupy over 46\% of $F_{tot}$. This influence is clearly seen in the shape of the radial distribution functions of floes $g_{rdf}$ (see Section~\ref{sec:theory}). For the initial and final states of model runs with $\bar{A}=0.6$ and $\bar{A}=0.9$, $g_{rdf}$ is shown in Fig.~\ref{fig:rdf}. As can be seen, the dominating feature of $g_{rdf}$ is a pattern of shallow minima separated with narrow spikes corresponding to the radii of the largest floes, i.e., reflecting small floes densely packed along their edges. Interestingly, for $\bar{A}=0.9$, the number of small floes surrounding a very large one is roughly proportional to its perimeter; hence an approximately linear dependence of the height of the peaks of $g_{rdf}(x)$ on $x$ in Fig.~\ref{fig:rdf}c.

The $g_{rdf}$ properties described above are present already in the initial, random floe distributions (in Fig.~\ref{fig:rdf}c the two curves are hardly distinguishable) and result simply from geometrical constraints of non-overlapping floes. Hence, $g_{rdf}$ is a poor indicator of the degree of clustering, especially for high ice concentrations (Fig.~\ref{fig:rdf}c). An alternative measure of clustering, suitable for very wide particle-size distributions, is therefore desirable. A straightforward candidate is a floe--floe distance distribution $g_{ffd}$, where the floe--floe distance is measured not between the floes' centers, but between their edges ($x=||\mathbf{x}_i-\mathbf{x}_j||-r_i-r_j$ instead of $x=||\mathbf{x}_i-\mathbf{x}_j||$, as in $g_{rdf}$). For random floe distribution in space, $g_{ffd}$ is approximately distance-independent at small ice concentrations (Fig.~\ref{fig:rdf}b) and slightly increases toward zero in a densely packed ice field, again because of the above-mentioned geometrical constraints (Fig.~\ref{fig:rdf}d). However, in a clustered state $g_{ffd}$ increases rapidly (up to an order of magnitude) towards zero.

In the next section, the value of $d_m=0.5$~m was used to define clusters (see Section~\ref{sec:theory}). Other tested values from the range 0.1--1.0~m gave very similar results.

\subsection{Cluster size distribution}\label{sec:resultsC}

A deeper insight into cluster formation and properties can be gained from an analysis of cluster-size distribution. Let $N_c$ denote the number of clusters in the analyzed snapshot, and $\mathcal{C}_k$ -- an $n_k$-element set of floes belonging to the $k$-th cluster ($k=1,\dots,N_c$ and $\sum_{k=1}^{N_c}n_k=N$). In the following, a total surface area of floes building the $k$-th cluster, $F_{c,k}$, is used as a measure of its size:
\begin{equation}\label{eq:Fc}
    F_{c,k} = {\pi\over4}\sum_{i\in\mathcal{C}_k}d_i^2.
\end{equation}
Because of generally very irregular shapes of clusters, especially at medium ice concentrations (Fig.~\ref{fig:snapshots}), this quantity seems more appropriate than e.g. the area of the sea surface occupied by a cluster, which is difficult to estimate reliably (in particular, a straightforward convex-hull approach turned out inappropriate). Obviously, $\sum_{k=1}^{N_c}F_{c,k}=F_{tot}=\bar{A}L^2$. The effective cluster radius is given by $r_{e,k} = (F_{c,k}/\pi)^{1/2}$.

Figure~\ref{fig:clustareadistrib} shows the rank-order distribution (see Appendix~\ref{sec:A_exponents}) of $r_e$ in the final, quasi-stationary state for three values of $\bar{A}$: 0.6, 0.75 and 0.9. The distributions have two dominating features. Firstly, over most of the range of values they are of power-law type:
\begin{equation}\label{eq:Pr_e}
     P(r_e;\bar{A})\sim r_e^{-1-\alpha(\bar{A})}.
\end{equation}
Secondly, the sizes of the largest clusters clearly deviate from the power-law regime, the more so the larger the ice concentration.

\begin{figure}
  \noindent\includegraphics[width=20pc]{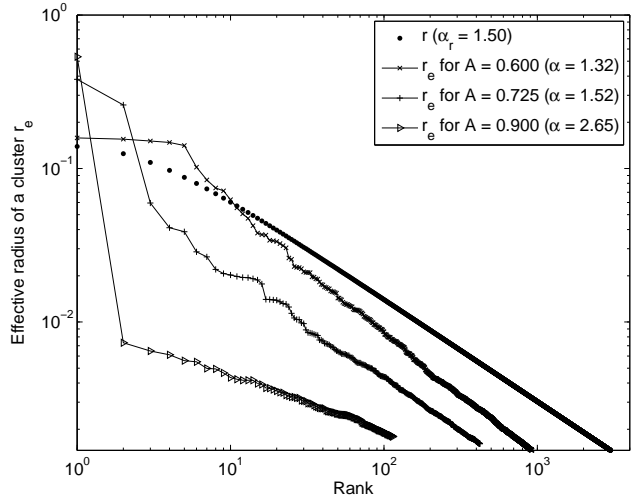}\\
  \caption{Rank-order distributions of effective cluster radius $r_e=(F_c/\pi)^{1/2}$ in fully clustered states for $\bar{A}$ equal 0.6, 0.725 and 0.9. Black dots show the corresponding distribution of the radii of single floes. Values are normalized so that $\pi\sum r_{e,i}^2=1$.}\label{fig:clustareadistrib}
\end{figure}

\begin{figure}
  \noindent\includegraphics[width=20pc]{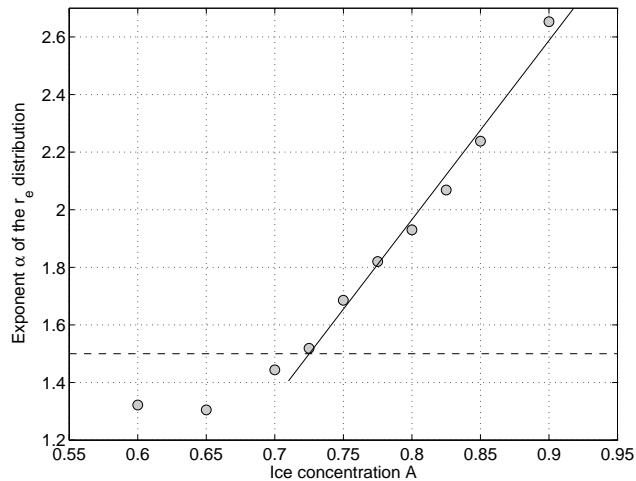}\\
  \caption{Exponent $\alpha$ of the distributions $P(r_e)$ of effective cluster radii $r_e$ (equation~\ref{eq:Pr_e}) in function of the mean ice concentration $\bar{A}$. The horizontal dashed line shows the value of $\alpha_r=1.5$. The continuous line shows the linear least-square fit to the data for $\bar{A}>0.7$. }\label{fig:mu_vs_A}
\end{figure}

The exponents $\alpha$ of probability density functions $P(r_e)$, estimated with a maximum likelihood method (equation~\ref{eqAp:alpha} in Appendix~\ref{sec:A_exponents}) increase with increasing $\bar{A}$, from values lower than $\alpha_r$ characteristic for low ice concentrations to values higher than $\alpha_r$ for densely packed ice fields (Fig.~\ref{fig:mu_vs_A}). The relationship between $\alpha_r$ and $\alpha$ suggests the existence of different mechanisms dominating cluster formation at low and high ice concentration. In the first case, typical clusters consist of the smallest floes accumulated along the downwind edges of the largest ones (Fig.~\ref{fig:snapshots}a), because of large differences between the respective equilibrium velocities. This effect is also present during the initial stages of cluster development at higher ice concentrations (not shown), similarly leading to a transient decrease of $\alpha$. In other words, the smallest floes tend to belong to the largest clusters, without contributing much to $F_c$, dominated by the area of the largest floes. Hence, for the largest clusters their $r_e$s are hardly larger than the respective radii of their largest members.

To the contrary, at high ice concentrations, due to high collision rates, the floes tend to move with similar translational velocity independently of their size (Fig.~\ref{fig:velcomps}c). In this case, one large-scale cluster develops, spanning the whole model domain (Figs.~\ref{fig:snapshots}c and \ref{fig:clustareadistrib}). Only the smallest floes can gain enough kinetic energy at collisions so that they are able to escape the mega-cluster; and only the smallest floes find enough empty space in its `holes' for an independent motion. This results in $\alpha>\alpha_r$. Finding a quantitative explanation for the observed dependence $\alpha(\bar{A})$, as shown in Fig.~\ref{fig:mu_vs_A}, is beyond the scope of this paper.

\begin{figure}[!t]
  \noindent\includegraphics[width=20pc]{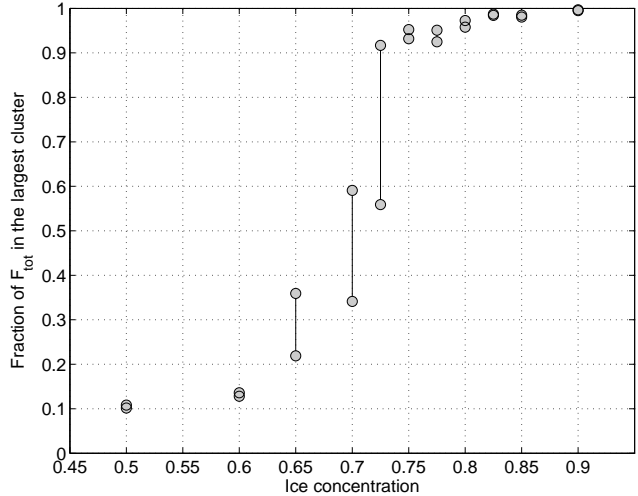}\\
  \caption{Surface area of the largest cluster $F_{c,1}$ (relative to the total surface area of all floes, $F_{tot}$) in function of the mean ice concentration $\bar{A}$. For each value of $\bar{A}$, the minimum and maximum values of $F_{c,1}$ recorded in the quasi-stationary state is shown.}\label{fig:Farea_vs_A}
\end{figure}

In terms of the dependence of the largest cluster size on the ice concentration, two regimes can be identified, as shown in Fig.~\ref{fig:Farea_vs_A}. For $\bar{A}$<$\sim$0.7, the area of the largest cluster remains comparable with the area of the largest floe in the sample (Fig.~\ref{fig:clustareadistrib}). For $\bar{A}\geq\sim0.75$, one dominating mega-cluster develops, occupying over 85\% of $F_{tot}$. Those two regimes are separated by a narrow range of medium, `critical' ice concentrations $\bar{A}_{crit}$, characterized by a very interesting dynamics. Contrary to a relatively smooth development of the largest cluster in time, observed for both $\bar{A}\ll\bar{A}_{crit}$ and $\bar{A}\gg\bar{A}_{crit}$, the transitional range is associated with very strong, erratic temporal variations of the size of the largest cluster, corresponding to its constant rearrangement, break-ups and re-consolidation. In a sense, when judged by the size of the largest cluster, the system never reaches a quasi-stationary state for $\bar{A}\simeq\bar{A}_{crit}$.

Interestingly, in this transitional state the exponents $\alpha$ characterizing the cluster-size distribution $P(r_e)$ are close to the value of $\alpha_r$ (Figs.~\ref{fig:clustareadistrib} and~\ref{fig:mu_vs_A}), i.e., the cluster-size distribution $P(r_e)$ mirrors $P(r)$. If $\alpha(\bar{A})$ is concerned, the critical range of $\bar{A}$ separates a region of slowly-varying values of $\alpha<\alpha_r$ from a region of fast, approximately linearly increasing values of $\alpha>\alpha_r$ (Fig.~\ref{fig:mu_vs_A}).

\section{Discussion and conclusions}\label{sec:conclusions}

The model presented in this paper describes a very idealized dynamics of MCIZ. Its purpose is rather to assist in understanding of basic processes regulating the formation of clusters in sea ice, than to reproduce details of any particular real-world situation. Qualitatively, the results of this study reproduce the dominant patterns of motion and clustering observed in MCIZ, including realistic granular temperature levels and clusters of small floes being pushed in front of a much larger one. Quantitative verification will require, on the one hand, experimental data on high-frequency sea-ice motion, unavailable at present, and on the other hand, more extensive simulations with a wide range of model parameters ($h$, $\mathbf{u}_a$, $\epsilon$, $\alpha_r$ etc.). Possible directions of future research include, but are not limited to: (i) formulation of an improved collisional rheology for MCIZ, based on internal-stress tensor components calculated with the SFC model, (ii) better understanding of the atmosphere--ice--ocean momentum transfer in MCIZ, resulting from the knowledge of the average motion of clustered ice fields in response to winds and currents, (iii) more detailed theoretical analysis of phenomena described in this paper, in particular explaining the dependence between the exponents of the floe-size and cluster-size distribution, and the nature of the transition between the two cluster-size regimes observed at low and high ice concentration, (iv) more detailed, in-depth analysis of the analogies between sea ice in MCIZ and other forced polydisperse granular media, and (v) analysis of the role of clustering in atmosphere--ocean heat transport and freezing/melting of the ice.

\begin{figure}
  \noindent\includegraphics[width=20pc]{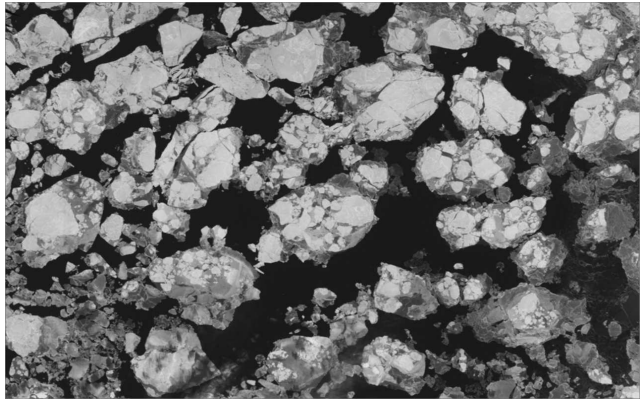}\\
  \caption{Fragment of a Landsat image of the Okhotsk Sea (3. Mar. 2003,
  157.1$^\circ$E, 58.7$^\circ$N) with clusters of thick ice floes
  consolidated by thin ice. Source: \cite{LandsatData}.}\label{fig:frozenclusters}
\end{figure}

The last point is particularly interesting in view of a hypothesis formulated recently by Toyota and colleagues \cite{Toyota11} who propose that cluster formation (or herding, in their nomenclature) may play a role in freezing of neighboring ice floes, and thus have an influence on the floe-size distribution in MCIZ. Combined with the results of the present study, this suggests an interesting possibility of a feedback between FSD and floe clustering. On the one hand, the FSD influences processes of cluster formation, including ice concentration within clusters and their size distribution. On the other hand, the existence of clusters may be important for floe formation in periods with low temperature, contributing to more intensive lateral freezing between densely packed floes. Clusters of floes frozen together are seen in satellite images of MCIZ (see Fig.~\ref{fig:frozenclusters} for an example of a Landsat image of the Okhotsk Sea). However, from a single snapshot it is not possible to determine the `life history' of those floe formations. They could have been produced by freezing within already existing clusters, or -- alternatively -- they could have been a result of a divergence of an initially densely packed, frozen floe field and of subsequent random breaking of thin ice occupying spaces between thick floes. In the second case, the end result could have formed without the clustering--freezing--FSD feedback proposed above. Estimating the plausibility of that feedback is not possible without more observational data.

Even though substantial progress has been made in recent years in observational techniques of the polar environment, including sea ice, MCIZ is still a very demanding, hard-to-explore medium. Consequently, even though it plays a crucial role in seasonal and long-term expansion and retreat of the ice cover in polar seas -- one of important proxies of the climate change -- the dynamics of MCIZ still remains poorly understood. Numerical models, like the one developed in this study, provide a valuable tool enabling to improve our knowledge and to gain more insight into functioning of that complex and fascinating environment.

\appendix

\section{Rank-ordering statistics and maximum-likelihood estimation}

To avoid drawbacks of binning in the analysis of distributions of the modeled quantities, rank-ordering technique is used throughout this paper \cite{Sornette06,Sornetteetal96}. It is based on data sorted in descending order and -- graphically -- plotted versus the ranks (as in Fig.~\ref{fig:clustareadistrib}).

\subsection{Calculation of the floe diameters}\label{sec:A_fsd}

It is assumed that the floe-size probability density function (pdf) $P(d)$ is given by a power law $P(d)=Cd^{-1-\alpha_r}$ with $d>0$ and $\alpha_r>1$, so that the mean $\bar{d}$ exists. For prescribed $\alpha_r$, $\bar{d}$ and the finite sample size $N$, acting as a constraint, the maximum-likelihood estimation gives the most probable value of the $n$th of $N$ elements as \citep[][p.~171]{Sornette06}:
\begin{equation}\label{eqAp:rmp}
    d_n = \left[{\alpha_r N+1\over\alpha n+1}C\right]^{1/\alpha_r}.
\end{equation}
The influence of the finite sample size is visible in the range of the largest elements in the sample: with decreasing $N$, the probability of encountering very large values decreases. This effect is seen in Fig. ~\ref{fig:clustareadistrib} as a deviation of the rank-order distribution of floe diameters from a straight line.

In the SFC model, formula~(\ref{eqAp:rmp}) is used with $C=1$, and the obtained values are subsequently re-scaled so that $\sum_{n=1}^N d_n = N\bar{d}$. For a given ice concentration $\bar{A}$, the length $L$ of the computational domain is calculated as $L=(\pi/(4\bar{A})\sum_{n=1}^Nd_n^2)^{1/2}$.

\subsection{Estimation of power-law exponents}\label{sec:A_exponents}

Let $P(x)$ denote a power-law probability density function of variable $X$, represented by $N$ rank-ordered data points $x_n$ ($n=1,\dots,N$). A maximum-likelihood estimation $\alpha_k$ of the exponent $\alpha$ of $P(x)$, based on the sub-sample $(x_k,x_N)$, where ($1\leq k<N$), is given by \cite{Sornette06}:
\begin{equation}\label{eqAp:alpha}
    \alpha_k = N\left[\sum_{n=k}^N\log{x_n\over x_N}\right]^{-1}.
\end{equation}
The standard deviation $\Delta\alpha_k$ of this estimate decreases with increasing sample size: $\Delta\alpha_k = \alpha_k(N-k+1)^{-1/2}$.


\begin{thebibliography}{38}%
\makeatletter
\providecommand \@ifxundefined [1]{%
 \@ifx{#1\undefined}
}%
\providecommand \@ifnum [1]{%
 \ifnum #1\expandafter \@firstoftwo
 \else \expandafter \@secondoftwo
 \fi
}%
\providecommand \@ifx [1]{%
 \ifx #1\expandafter \@firstoftwo
 \else \expandafter \@secondoftwo
 \fi
}%
\providecommand \natexlab [1]{#1}%
\providecommand \enquote  [1]{``#1''}%
\providecommand \bibnamefont  [1]{#1}%
\providecommand \bibfnamefont [1]{#1}%
\providecommand \citenamefont [1]{#1}%
\providecommand \href@noop [0]{\@secondoftwo}%
\providecommand \href [0]{\begingroup \@sanitize@url \@href}%
\providecommand \@href[1]{\@@startlink{#1}\@@href}%
\providecommand \@@href[1]{\endgroup#1\@@endlink}%
\providecommand \@sanitize@url [0]{\catcode `\\12\catcode `\$12\catcode
  `\&12\catcode `\#12\catcode `\^12\catcode `\_12\catcode `\%12\relax}%
\providecommand \@@startlink[1]{}%
\providecommand \@@endlink[0]{}%
\providecommand \url  [0]{\begingroup\@sanitize@url \@url }%
\providecommand \@url [1]{\endgroup\@href {#1}{\urlprefix }}%
\providecommand \urlprefix  [0]{URL }%
\providecommand \Eprint [0]{\href }%
\providecommand \doibase [0]{http://dx.doi.org/}%
\providecommand \selectlanguage [0]{\@gobble}%
\providecommand \bibinfo  [0]{\@secondoftwo}%
\providecommand \bibfield  [0]{\@secondoftwo}%
\providecommand \translation [1]{[#1]}%
\providecommand \BibitemOpen [0]{}%
\providecommand \bibitemStop [0]{}%
\providecommand \bibitemNoStop [0]{.\EOS\space}%
\providecommand \EOS [0]{\spacefactor3000\relax}%
\providecommand \BibitemShut  [1]{\csname bibitem#1\endcsname}%
\let\auto@bib@innerbib\@empty
\bibitem [{\citenamefont {Serreze}\ \emph {et~al.}(2007)\citenamefont
  {Serreze}, \citenamefont {Holland},\ and\ \citenamefont
  {Stroeve}}]{Serrezeetal07}%
  \BibitemOpen
  \bibfield  {author} {\bibinfo {author} {\bibfnamefont {M.~C.}\ \bibnamefont
  {Serreze}}, \bibinfo {author} {\bibfnamefont {M.~M.}\ \bibnamefont
  {Holland}}, \ and\ \bibinfo {author} {\bibfnamefont {J.}~\bibnamefont
  {Stroeve}},\ }\href@noop {} {\bibfield  {journal} {\bibinfo  {journal}
  {Science}\ }\textbf {\bibinfo {volume} {315}},\ \bibinfo {pages} {1533}
  (\bibinfo {year} {2007})},\ \bibinfo {note}
  {doi:10.1126/science.1139426}\BibitemShut {NoStop}%
\bibitem [{\citenamefont {Maslanik}\ \emph {et~al.}(2007)\citenamefont
  {Maslanik}, \citenamefont {Fowler}, \citenamefont {Stroeve}, \citenamefont
  {Drobot}, \citenamefont {Zwally}, \citenamefont {Yi},\ and\ \citenamefont
  {Emery}}]{Maslaniketal07}%
  \BibitemOpen
  \bibfield  {author} {\bibinfo {author} {\bibfnamefont {J.~A.}\ \bibnamefont
  {Maslanik}}, \bibinfo {author} {\bibfnamefont {C.}~\bibnamefont {Fowler}},
  \bibinfo {author} {\bibfnamefont {J.}~\bibnamefont {Stroeve}}, \bibinfo
  {author} {\bibfnamefont {S.}~\bibnamefont {Drobot}}, \bibinfo {author}
  {\bibfnamefont {J.}~\bibnamefont {Zwally}}, \bibinfo {author} {\bibfnamefont
  {D.}~\bibnamefont {Yi}}, \ and\ \bibinfo {author} {\bibfnamefont
  {W.}~\bibnamefont {Emery}},\ }\href@noop {} {\bibfield  {journal} {\bibinfo
  {journal} {Geophys. Res. Lett.}\ }\textbf {\bibinfo {volume} {34}},\ \bibinfo
  {pages} {L24501} (\bibinfo {year} {2007})},\ \bibinfo {note}
  {doi:10.1029/2007GL032043}\BibitemShut {NoStop}%
\bibitem [{\citenamefont {Rothrock}\ \emph {et~al.}(2008)\citenamefont
  {Rothrock}, \citenamefont {Percival},\ and\ \citenamefont
  {Wensnahan}}]{Rothrocketal08}%
  \BibitemOpen
  \bibfield  {author} {\bibinfo {author} {\bibfnamefont {D.~A.}\ \bibnamefont
  {Rothrock}}, \bibinfo {author} {\bibfnamefont {D.~B.}\ \bibnamefont
  {Percival}}, \ and\ \bibinfo {author} {\bibfnamefont {M.}~\bibnamefont
  {Wensnahan}},\ }\href@noop {} {\bibfield  {journal} {\bibinfo  {journal} {J.
  Geophys. Res.}\ }\textbf {\bibinfo {volume} {113}},\ \bibinfo {pages}
  {C05003} (\bibinfo {year} {2008})},\ \bibinfo {note}
  {doi:10.1029/2007JC004252}\BibitemShut {NoStop}%
\bibitem [{\citenamefont {Serreze}(2011)}]{Serreze11}%
  \BibitemOpen
  \bibfield  {author} {\bibinfo {author} {\bibfnamefont {M.~C.}\ \bibnamefont
  {Serreze}},\ }\href@noop {} {\bibfield  {journal} {\bibinfo  {journal}
  {Nature}\ }\textbf {\bibinfo {volume} {471}},\ \bibinfo {pages} {47}
  (\bibinfo {year} {2011})},\ \bibinfo {note} {doi:10.1038/471047a}\BibitemShut
  {NoStop}%
\bibitem [{\citenamefont {Aranson}\ and\ \citenamefont
  {Tsimring}(2006)}]{AransonTsimring06}%
  \BibitemOpen
  \bibfield  {author} {\bibinfo {author} {\bibfnamefont {I.~S.}\ \bibnamefont
  {Aranson}}\ and\ \bibinfo {author} {\bibfnamefont {L.~S.}\ \bibnamefont
  {Tsimring}},\ }\href@noop {} {\bibfield  {journal} {\bibinfo  {journal} {Rev.
  Modern Phys.}\ }\textbf {\bibinfo {volume} {78}},\ \bibinfo {pages} {641}
  (\bibinfo {year} {2006})}\BibitemShut {NoStop}%
\bibitem [{\citenamefont {Shen}\ \emph {et~al.}(1984)\citenamefont {Shen},
  \citenamefont {\hbox{Hibler~III}},\ and\ \citenamefont
  {Lepp\"aranta}}]{Shen84}%
  \BibitemOpen
  \bibfield  {author} {\bibinfo {author} {\bibfnamefont {H.}~\bibnamefont
  {Shen}}, \bibinfo {author} {\bibfnamefont {W.}~\bibnamefont
  {\hbox{Hibler~III}}}, \ and\ \bibinfo {author} {\bibfnamefont
  {M.}~\bibnamefont {Lepp\"aranta}},\ }\href@noop {} {\bibfield  {journal}
  {\bibinfo  {journal} {MIZEX Bulletin III, USACREL Special Report 84-28}\ ,\
  \bibinfo {pages} {29}} (\bibinfo {year} {1984})}\BibitemShut {NoStop}%
\bibitem [{\citenamefont {Shen}\ \emph {et~al.}(1986)\citenamefont {Shen},
  \citenamefont {\hbox{Hibler~III}},\ and\ \citenamefont
  {Lepp\"aranta}}]{Shen86}%
  \BibitemOpen
  \bibfield  {author} {\bibinfo {author} {\bibfnamefont {H.}~\bibnamefont
  {Shen}}, \bibinfo {author} {\bibfnamefont {W.}~\bibnamefont
  {\hbox{Hibler~III}}}, \ and\ \bibinfo {author} {\bibfnamefont
  {M.}~\bibnamefont {Lepp\"aranta}},\ }\href@noop {} {\bibfield  {journal}
  {\bibinfo  {journal} {Acta Mechanica}\ }\textbf {\bibinfo {volume} {63}},\
  \bibinfo {pages} {143} (\bibinfo {year} {1986})}\BibitemShut {NoStop}%
\bibitem [{\citenamefont {Lu}\ \emph {et~al.}(1989)\citenamefont {Lu},
  \citenamefont {Larsen},\ and\ \citenamefont {Tryde}}]{Lu89}%
  \BibitemOpen
  \bibfield  {author} {\bibinfo {author} {\bibfnamefont {Q.}~\bibnamefont
  {Lu}}, \bibinfo {author} {\bibfnamefont {J.}~\bibnamefont {Larsen}}, \ and\
  \bibinfo {author} {\bibfnamefont {P.}~\bibnamefont {Tryde}},\ }\href@noop {}
  {\bibfield  {journal} {\bibinfo  {journal} {J. Geophys. Res.}\ }\textbf
  {\bibinfo {volume} {94}},\ \bibinfo {pages} {14525} (\bibinfo {year}
  {1989})}\BibitemShut {NoStop}%
\bibitem [{\citenamefont {Feltham}(2005)}]{Feltham05}%
  \BibitemOpen
  \bibfield  {author} {\bibinfo {author} {\bibfnamefont {D.}~\bibnamefont
  {Feltham}},\ }\href@noop {} {\bibfield  {journal} {\bibinfo  {journal} {Phyl.
  Trans. Royal Soc. A}\ }\textbf {\bibinfo {volume} {363}},\ \bibinfo {pages}
  {1677} (\bibinfo {year} {2005})},\ \bibinfo {note}
  {doi:10.1098/rsta.2005.1601}\BibitemShut {NoStop}%
\bibitem [{\citenamefont {Goldhirsch}\ and\ \citenamefont
  {Zanetti}(1993)}]{GoldhirschZanetti93}%
  \BibitemOpen
  \bibfield  {author} {\bibinfo {author} {\bibfnamefont {I.}~\bibnamefont
  {Goldhirsch}}\ and\ \bibinfo {author} {\bibfnamefont {G.}~\bibnamefont
  {Zanetti}},\ }\href@noop {} {\bibfield  {journal} {\bibinfo  {journal} {Phys.
  Rev. Lett.}\ }\textbf {\bibinfo {volume} {70}},\ \bibinfo {pages} {1619}
  (\bibinfo {year} {1993})}\BibitemShut {NoStop}%
\bibitem [{\citenamefont {Brey}\ and\ \citenamefont
  {Cubero}(2001)}]{BreyCubero01}%
  \BibitemOpen
  \bibfield  {author} {\bibinfo {author} {\bibfnamefont {J.~J.}\ \bibnamefont
  {Brey}}\ and\ \bibinfo {author} {\bibfnamefont {D.}~\bibnamefont {Cubero}},\
  }in\ \href@noop {} {\emph {\bibinfo {booktitle} {Granular Gases}}},\ \bibinfo
  {series} {Lecture Notes in Physics}, Vol.\ \bibinfo {volume} {564},\ \bibinfo
  {editor} {edited by\ \bibinfo {editor} {\bibfnamefont {T.}~\bibnamefont
  {P\"oschel}}\ and\ \bibinfo {editor} {\bibfnamefont {S.}~\bibnamefont
  {Luding}}}\ (\bibinfo  {publisher} {Springer, Berlin--Heidelberg},\ \bibinfo
  {year} {2001})\ pp.\ \bibinfo {pages} {59--78}\BibitemShut {NoStop}%
\bibitem [{Lan(2011)}]{LandsatData}%
  \BibitemOpen
  \href@noop {} {\bibinfo {title} {U.S. Geological Survey},}\ 
  \bibinfo {note}
  {http://glovis.usgs.gov/}\ (\bibinfo {year} {2011})\BibitemShut {NoStop}%
\bibitem [{\citenamefont {Inoue}\ \emph {et~al.}(2004)\citenamefont {Inoue},
  \citenamefont {Wakatsuchi},\ and\ \citenamefont {Fujiyoshi}}]{Inoue04}%
  \BibitemOpen
  \bibfield  {author} {\bibinfo {author} {\bibfnamefont {J.}~\bibnamefont
  {Inoue}}, \bibinfo {author} {\bibfnamefont {M.}~\bibnamefont {Wakatsuchi}}, \
  and\ \bibinfo {author} {\bibfnamefont {Y.}~\bibnamefont {Fujiyoshi}},\
  }\href@noop {} {\bibfield  {journal} {\bibinfo  {journal} {Geophys. Res.
  Lett.}\ }\textbf {\bibinfo {volume} {31}},\ \bibinfo {pages} {L20303}
  (\bibinfo {year} {2004})},\ \bibinfo {note}
  {doi:10.1029/2004GL020809}\BibitemShut {NoStop}%
\bibitem [{\citenamefont {Toyota}\ \emph {et~al.}(2006)\citenamefont {Toyota},
  \citenamefont {Takatsuji},\ and\ \citenamefont {Nakayama}}]{Toyota06}%
  \BibitemOpen
  \bibfield  {author} {\bibinfo {author} {\bibfnamefont {T.}~\bibnamefont
  {Toyota}}, \bibinfo {author} {\bibfnamefont {S.}~\bibnamefont {Takatsuji}}, \
  and\ \bibinfo {author} {\bibfnamefont {M.}~\bibnamefont {Nakayama}},\
  }\href@noop {} {\bibfield  {journal} {\bibinfo  {journal} {Geophys. Res.
  Lett.}\ }\textbf {\bibinfo {volume} {33}},\ \bibinfo {pages} {L02616}
  (\bibinfo {year} {2006})},\ \bibinfo {note}
  {doi:10.1029/2005GL024556}\BibitemShut {NoStop}%
\bibitem [{\citenamefont {Steer}\ \emph {et~al.}(2008)\citenamefont {Steer},
  \citenamefont {Worby},\ and\ \citenamefont {Heil}}]{Steer08}%
  \BibitemOpen
  \bibfield  {author} {\bibinfo {author} {\bibfnamefont {A.}~\bibnamefont
  {Steer}}, \bibinfo {author} {\bibfnamefont {A.}~\bibnamefont {Worby}}, \ and\
  \bibinfo {author} {\bibfnamefont {P.}~\bibnamefont {Heil}},\ }\href@noop {}
  {\bibfield  {journal} {\bibinfo  {journal} {Deep-Sea Res.~II}\ }\textbf
  {\bibinfo {volume} {55}},\ \bibinfo {pages} {933} (\bibinfo {year} {2008})},\
  \bibinfo {note} {doi:10.1016/j.dsr2.2007.12.016}\BibitemShut {NoStop}%
\bibitem [{\citenamefont {Herman}(2010)}]{Herman10}%
  \BibitemOpen
  \bibfield  {author} {\bibinfo {author} {\bibfnamefont {A.}~\bibnamefont
  {Herman}},\ }\href@noop {} {\bibfield  {journal} {\bibinfo  {journal} {Phys.
  Rev. E}\ }\textbf {\bibinfo {volume} {81}},\ \bibinfo {pages} {066123}
  (\bibinfo {year} {2010})}\BibitemShut {NoStop}%
\bibitem [{\citenamefont {Toyota}\ \emph {et~al.}(2011)\citenamefont {Toyota},
  \citenamefont {Haas},\ and\ \citenamefont {Tamura}}]{Toyota11}%
  \BibitemOpen
  \bibfield  {author} {\bibinfo {author} {\bibfnamefont {T.}~\bibnamefont
  {Toyota}}, \bibinfo {author} {\bibfnamefont {C.}~\bibnamefont {Haas}}, \ and\
  \bibinfo {author} {\bibfnamefont {T.}~\bibnamefont {Tamura}},\ }\href@noop {}
  {\bibfield  {journal} {\bibinfo  {journal} {Deep Sea Res. II}\ }\textbf
  {\bibinfo {volume} {9--10}},\ \bibinfo {pages} {1182} (\bibinfo {year}
  {2011})},\ \bibinfo {note} {doi:10.1016/j.dsr2.2010.10.034}\BibitemShut
  {NoStop}%
\bibitem [{\citenamefont {Hunke}\ and\ \citenamefont
  {Lipscomb}(2010)}]{CICE41}%
  \BibitemOpen
  \bibfield  {author} {\bibinfo {author} {\bibfnamefont {E.}~\bibnamefont
  {Hunke}}\ and\ \bibinfo {author} {\bibfnamefont {W.}~\bibnamefont
  {Lipscomb}},\ }\href@noop {} {\emph {\bibinfo {title} {C{I}{C}{E}: the {L}os
  {A}lamos {S}ea {I}ce {M}odel {D}ocumentation and {S}oftware {U}ser's {M}anual
  {V}ersion 4.1}}},\ \bibinfo {type} {Tech. Rep.}\ \bibinfo {number}
  {LA-CC-06-012}\ (\bibinfo  {institution} {Los Alamos National Laboratory},\
  \bibinfo {year} {2010})\ \bibinfo {note} {76~pp.}\BibitemShut {Stop}%
\bibitem [{\citenamefont {Luding}(2009)}]{Luding09}%
  \BibitemOpen
  \bibfield  {author} {\bibinfo {author} {\bibfnamefont {S.}~\bibnamefont
  {Luding}},\ }\href@noop {} {\bibfield  {journal} {\bibinfo  {journal}
  {Nonlinearity}\ }\textbf {\bibinfo {volume} {22}},\ \bibinfo {pages} {R101}
  (\bibinfo {year} {2009})}\BibitemShut {NoStop}%
\bibitem [{\citenamefont {Luding}\ and\ \citenamefont
  {Strau\ss}(2001)}]{LudingStrauss01}%
  \BibitemOpen
  \bibfield  {author} {\bibinfo {author} {\bibfnamefont {S.}~\bibnamefont
  {Luding}}\ and\ \bibinfo {author} {\bibfnamefont {O.}~\bibnamefont
  {Strau\ss}},\ }in\ \href@noop {} {\emph {\bibinfo {booktitle} {Granular
  Gases}}},\ \bibinfo {series} {Lecture Notes in Physics}, Vol.\ \bibinfo
  {volume} {564},\ \bibinfo {editor} {edited by\ \bibinfo {editor}
  {\bibfnamefont {T.}~\bibnamefont {P\"oschel}}\ and\ \bibinfo {editor}
  {\bibfnamefont {S.}~\bibnamefont {Luding}}}\ (\bibinfo  {publisher}
  {Springer, Berlin--Heidelberg},\ \bibinfo {year} {2001})\ pp.\ \bibinfo
  {pages} {389--409}\BibitemShut {NoStop}%
\bibitem [{\citenamefont {Dahl}\ \emph {et~al.}(2002)\citenamefont {Dahl},
  \citenamefont {Clelland},\ and\ \citenamefont
  {Hrenya}}]{DahlClellandHrenya2002}%
  \BibitemOpen
  \bibfield  {author} {\bibinfo {author} {\bibfnamefont {S.~R.}\ \bibnamefont
  {Dahl}}, \bibinfo {author} {\bibfnamefont {R.}~\bibnamefont {Clelland}}, \
  and\ \bibinfo {author} {\bibfnamefont {C.~M.}\ \bibnamefont {Hrenya}},\
  }\href@noop {} {\bibfield  {journal} {\bibinfo  {journal} {Phys. Fluids}\
  }\textbf {\bibinfo {volume} {14}},\ \bibinfo {pages} {1972} (\bibinfo {year}
  {2002})}\BibitemShut {NoStop}%
\bibitem [{\citenamefont {Rice}\ and\ \citenamefont
  {Hrenya}(2010)}]{RiceHrenya2010}%
  \BibitemOpen
  \bibfield  {author} {\bibinfo {author} {\bibfnamefont {R.~B.}\ \bibnamefont
  {Rice}}\ and\ \bibinfo {author} {\bibfnamefont {C.~M.}\ \bibnamefont
  {Hrenya}},\ }\href@noop {} {\bibfield  {journal} {\bibinfo  {journal} {Phys.
  Rev. E}\ }\textbf {\bibinfo {volume} {81}},\ \bibinfo {pages} {021302}
  (\bibinfo {year} {2010})}\BibitemShut {NoStop}%
\bibitem [{\citenamefont {Luding}\ and\ \citenamefont
  {Herrmann}(1999)}]{LudingHerrmann99}%
  \BibitemOpen
  \bibfield  {author} {\bibinfo {author} {\bibfnamefont {S.}~\bibnamefont
  {Luding}}\ and\ \bibinfo {author} {\bibfnamefont {H.~J.}\ \bibnamefont
  {Herrmann}},\ }\href@noop {} {\bibfield  {journal} {\bibinfo  {journal}
  {Chaos}\ }\textbf {\bibinfo {volume} {9}},\ \bibinfo {pages} {673} (\bibinfo
  {year} {1999})}\BibitemShut {NoStop}%
\bibitem [{\citenamefont {Luding}(2002)}]{Luding02}%
  \BibitemOpen
  \bibfield  {author} {\bibinfo {author} {\bibfnamefont {S.}~\bibnamefont
  {Luding}},\ }\href@noop {} {\bibfield  {journal} {\bibinfo  {journal} {C. R.
  Physique}\ }\textbf {\bibinfo {volume} {3}},\ \bibinfo {pages} {153}
  (\bibinfo {year} {2002})}\BibitemShut {NoStop}%
\bibitem [{\citenamefont {Miller}\ and\ \citenamefont
  {Luding}(2004)}]{MillerLuding04}%
  \BibitemOpen
  \bibfield  {author} {\bibinfo {author} {\bibfnamefont {S.}~\bibnamefont
  {Miller}}\ and\ \bibinfo {author} {\bibfnamefont {S.}~\bibnamefont
  {Luding}},\ }\href@noop {} {\bibfield  {journal} {\bibinfo  {journal} {Phys.
  Rev. E}\ }\textbf {\bibinfo {volume} {69}},\ \bibinfo {pages} {031305}
  (\bibinfo {year} {2004})}\BibitemShut {NoStop}%
\bibitem [{\citenamefont {Luding}(2005)}]{Luding05}%
  \BibitemOpen
  \bibfield  {author} {\bibinfo {author} {\bibfnamefont {S.}~\bibnamefont
  {Luding}},\ }\href@noop {} {\bibfield  {journal} {\bibinfo  {journal}
  {Pramana J. Physics}\ }\textbf {\bibinfo {volume} {64}},\ \bibinfo {pages}
  {893} (\bibinfo {year} {2005})}\BibitemShut {NoStop}%
\bibitem [{\citenamefont {Rice}\ and\ \citenamefont
  {Hrenya}(2009)}]{RiceHrenya2009}%
  \BibitemOpen
  \bibfield  {author} {\bibinfo {author} {\bibfnamefont {R.~B.}\ \bibnamefont
  {Rice}}\ and\ \bibinfo {author} {\bibfnamefont {C.~M.}\ \bibnamefont
  {Hrenya}},\ }\href@noop {} {\bibfield  {journal} {\bibinfo  {journal} {Phys.
  Rev. E}\ }\textbf {\bibinfo {volume} {79}},\ \bibinfo {pages} {021304}
  (\bibinfo {year} {2009})}\BibitemShut {NoStop}%
\bibitem [{\citenamefont {Allen}\ and\ \citenamefont
  {Tildesley}(1989)}]{AllenTildesley89}%
  \BibitemOpen
  \bibfield  {author} {\bibinfo {author} {\bibfnamefont {M.~P.}\ \bibnamefont
  {Allen}}\ and\ \bibinfo {author} {\bibfnamefont {D.~J.}\ \bibnamefont
  {Tildesley}},\ }\href@noop {} {\emph {\bibinfo {title} {Computer {S}imulation
  of {L}iquids}}}\ (\bibinfo  {publisher} {Oxford University Press},\ \bibinfo
  {year} {1989})\ \bibinfo {note} {408~pp.}\BibitemShut {Stop}%
\bibitem [{\citenamefont {Sigurgeirsson}\ \emph {et~al.}(2001)\citenamefont
  {Sigurgeirsson}, \citenamefont {Stuart},\ and\ \citenamefont
  {Wan}}]{Sigurgeirsson01}%
  \BibitemOpen
  \bibfield  {author} {\bibinfo {author} {\bibfnamefont {H.}~\bibnamefont
  {Sigurgeirsson}}, \bibinfo {author} {\bibfnamefont {A.}~\bibnamefont
  {Stuart}}, \ and\ \bibinfo {author} {\bibfnamefont {W.-L.}\ \bibnamefont
  {Wan}},\ }\href@noop {} {\bibfield  {journal} {\bibinfo  {journal} {J.
  Comput. Phys.}\ }\textbf {\bibinfo {volume} {172}},\ \bibinfo {pages} {766}
  (\bibinfo {year} {2001})}\BibitemShut {NoStop}%
\bibitem [{\citenamefont {Mar\'in}(1997)}]{Marin97}%
  \BibitemOpen
  \bibfield  {author} {\bibinfo {author} {\bibfnamefont {M.}~\bibnamefont
  {Mar\'in}},\ }\href@noop {} {\bibfield  {journal} {\bibinfo  {journal}
  {Computer Physics Communications}\ }\textbf {\bibinfo {volume} {102}},\
  \bibinfo {pages} {81} (\bibinfo {year} {1997})}\BibitemShut {NoStop}%
\bibitem [{\citenamefont {Miller}\ and\ \citenamefont
  {Luding}(2003)}]{MillerLuding03}%
  \BibitemOpen
  \bibfield  {author} {\bibinfo {author} {\bibfnamefont {S.}~\bibnamefont
  {Miller}}\ and\ \bibinfo {author} {\bibfnamefont {S.}~\bibnamefont
  {Luding}},\ }\href@noop {} {\bibfield  {journal} {\bibinfo  {journal} {J.
  Comp. Phys.}\ }\textbf {\bibinfo {volume} {193}},\ \bibinfo {pages} {306}
  (\bibinfo {year} {2003})}\BibitemShut {NoStop}%
\bibitem [{\citenamefont {Lepp\"aranta}(2005)}]{LepparantaBook}%
  \BibitemOpen
  \bibfield  {author} {\bibinfo {author} {\bibfnamefont {M.}~\bibnamefont
  {Lepp\"aranta}},\ }\href@noop {} {\emph {\bibinfo {title} {The {D}rift of
  {S}ea {I}ce}}}\ (\bibinfo  {publisher} {Springer-Verlag,
  Berlin--Heidelberg--New York},\ \bibinfo {year} {2005})\ \bibinfo {note}
  {267~pp.}\BibitemShut {Stop}%
\bibitem [{\citenamefont {Mai}\ \emph {et~al.}(1996)\citenamefont {Mai},
  \citenamefont {Wamser},\ and\ \citenamefont
  {Kottmeier}}]{MaiWamserKottmeier96}%
  \BibitemOpen
  \bibfield  {author} {\bibinfo {author} {\bibfnamefont {S.}~\bibnamefont
  {Mai}}, \bibinfo {author} {\bibfnamefont {C.}~\bibnamefont {Wamser}}, \ and\
  \bibinfo {author} {\bibfnamefont {C.}~\bibnamefont {Kottmeier}},\ }\href@noop
  {} {\bibfield  {journal} {\bibinfo  {journal} {Bound. Layer Meteorol.}\
  }\textbf {\bibinfo {volume} {77}},\ \bibinfo {pages} {233} (\bibinfo {year}
  {1996})}\BibitemShut {NoStop}%
\bibitem [{\citenamefont {Garbrecht}\ \emph {et~al.}(2002)\citenamefont
  {Garbrecht}, \citenamefont {L\"upkes}, \citenamefont {Hartmann},\ and\
  \citenamefont {Wolff}}]{Garbrechtetal02}%
  \BibitemOpen
  \bibfield  {author} {\bibinfo {author} {\bibfnamefont {T.}~\bibnamefont
  {Garbrecht}}, \bibinfo {author} {\bibfnamefont {C.}~\bibnamefont {L\"upkes}},
  \bibinfo {author} {\bibfnamefont {J.}~\bibnamefont {Hartmann}}, \ and\
  \bibinfo {author} {\bibfnamefont {M.}~\bibnamefont {Wolff}},\ }\href@noop {}
  {\bibfield  {journal} {\bibinfo  {journal} {Tellus}\ }\textbf {\bibinfo
  {volume} {54A}},\ \bibinfo {pages} {205} (\bibinfo {year}
  {2002})}\BibitemShut {NoStop}%
\bibitem [{\citenamefont {L\"upkes}\ and\ \citenamefont
  {Birnbaum}(2005)}]{LupkesBirnbaum05}%
  \BibitemOpen
  \bibfield  {author} {\bibinfo {author} {\bibfnamefont {C.}~\bibnamefont
  {L\"upkes}}\ and\ \bibinfo {author} {\bibfnamefont {G.}~\bibnamefont
  {Birnbaum}},\ }\href@noop {} {\bibfield  {journal} {\bibinfo  {journal}
  {Bound. Layer Meteorol.}\ }\textbf {\bibinfo {volume} {117}},\ \bibinfo
  {pages} {179} (\bibinfo {year} {2005})}\BibitemShut {NoStop}%
\bibitem [{\citenamefont {Luding}\ and\ \citenamefont
  {McNamara}(1998)}]{LudingMcNamara98}%
  \BibitemOpen
  \bibfield  {author} {\bibinfo {author} {\bibfnamefont {S.}~\bibnamefont
  {Luding}}\ and\ \bibinfo {author} {\bibfnamefont {S.}~\bibnamefont
  {McNamara}},\ }\href@noop {} {\bibfield  {journal} {\bibinfo  {journal}
  {Granular Matter}\ }\textbf {\bibinfo {volume} {1}},\ \bibinfo {pages} {113}
  (\bibinfo {year} {1998})}\BibitemShut {NoStop}%
\bibitem [{\citenamefont {Sornette}(2006)}]{Sornette06}%
  \BibitemOpen
  \bibfield  {author} {\bibinfo {author} {\bibfnamefont {D.}~\bibnamefont
  {Sornette}},\ }\href@noop {} {\emph {\bibinfo {title} {Critical {P}henomena
  in {N}atural {S}ciences}}},\ \bibinfo {edition} {2nd}\ ed.\ (\bibinfo
  {publisher} {Springer Berlin--Heidelberg},\ \bibinfo {year}
  {2006})\BibitemShut {NoStop}%
\bibitem [{\citenamefont {Sornette}\ \emph {et~al.}(1996)\citenamefont
  {Sornette}, \citenamefont {Knopoff}, \citenamefont {Kagan},\ and\
  \citenamefont {Vanneste}}]{Sornetteetal96}%
  \BibitemOpen
  \bibfield  {author} {\bibinfo {author} {\bibfnamefont {D.}~\bibnamefont
  {Sornette}}, \bibinfo {author} {\bibfnamefont {L.}~\bibnamefont {Knopoff}},
  \bibinfo {author} {\bibfnamefont {Y.~Y.}\ \bibnamefont {Kagan}}, \ and\
  \bibinfo {author} {\bibfnamefont {C.}~\bibnamefont {Vanneste}},\ }\href@noop
  {} {\bibfield  {journal} {\bibinfo  {journal} {J. Geophys. Res.}\ }\textbf
  {\bibinfo {volume} {101B}},\ \bibinfo {pages} {13883} (\bibinfo {year}
  {1996})}\BibitemShut {NoStop}%
\end{thebibliography}
%

\end{document}